\documentclass[conference]{IEEEtran}
\IEEEoverridecommandlockouts

\usepackage[utf8]{inputenc}
\usepackage[english]{babel}
\usepackage{csquotes,xpatch}
\usepackage[backend=biber,style=ieee]{biblatex}
\usepackage{booktabs}
\usepackage{listings}
\usepackage{amsmath,amssymb,amsfonts}
\usepackage{algorithmic}
\usepackage{graphicx}
\usepackage{textcomp}
\usepackage{xcolor}
\usepackage{cleveref}

\def\BibTeX{{\rm B\kern-.05em{\sc i\kern-.025em b}\kern-.08em
    T\kern-.1667em\lower.7ex\hbox{E}\kern-.125emX}}
\begin{document}

\title{Building an ID Card Repository with Progressive Web Application to Mitigate Fraud}

\author{\IEEEauthorblockN{1\textsuperscript{st} Kevin Akbar Adhiguna}
\IEEEauthorblockA{\textit{Faculty of Mathematics and Natural Science} \\
\textit{Padjadjaran University}\\
Sumedang district, West Java, Indonesia \\
kevin17016@mail.unpad.ac.id}
\and
\IEEEauthorblockN{2\textsuperscript{nd} Firhan Maulana Rusli}
\IEEEauthorblockA{\textit{School of Computing} \\
\textit{Telkom University}\\
Bandung City, West Java, Indonesia \\
firhanmaulanarusli@gmail.com}
\and
\IEEEauthorblockN{3\textsuperscript{rd} Hendy Irawan}
\IEEEauthorblockA{\textit{School of Computing} \\
\textit{Telkom University}\\
Bandung City, West Java, Indonesia \\
hendy@lovia.life}
}

\maketitle

\begin{abstract}
A lot of service requires identity of users to mitigate undesirable incidents, such as fraud. To cut down probability of potential fraud, ID Card of users are collected to be verified so people can verify users whenever an undesirable activity happens. However, to verify a person’s identity through his/her ID Card, a repository for the ID Card is required. To verify ID cards, the ID Card repository will be connected to an automated ID Card Verification API. The ID Card repository is meant to be used both on mobile phones and desktop computers, so the concept of progressive web application is used. To be able to upload images smoothly and build progressive web application, the ID Card repository is built using ReactJS and Ant Design. Server side is powered by Strapi and MongoDB. GraphQL API is utilized to connect client side and server side. It involves queries to fetch data. To fetch data on the client side, Apollo client is used to  in ReactJS. Git is utilized for version control system which gives contribution to Continuous Integration/Continuous Delivery (CI/CD). In this paper, we will discuss why Strapi is best suited on server side and how Ant Design, a library to style components in a web page, can provide required components in this web application.
\end{abstract}


\section{Introduction}
Data and information have been a part of human life. They are mostly accessed through the Internet and can be provided in various ways such as text, files, and images. People upload and download data. However, when developing a web application that enables its users to register, enterprises might want to verify a user's identity. One common way is to ask users to upload his/her identity card. ID card repository will be used to save user's ID card.

User Interface (UI) is important in creating an application, whether it is a mobile application or a web application. However, creating application's functionality is a lot more important than UI and those things are often time-consuming. Choosing an appropriate UI library becomes vital as necessary web application components must be provided.

On the server side, there are notable things that must be considered seriously. The first one is Application Programming Interface (API) creation. Other than Simple Object Access Protocol (SOAP), Restful API (REST API) and GraphQL API are available \autocite{katayama2010togows, nogatz2017implementing}. Both REST API and GraphQL API have drawbacks and advantages, so it is desirable to have a library or framework that supports both. In order to focus on the web application's functionality, API creation time and effort need to be reduced.

Another notable thing is related to security which are authentication and authorization. JSON Web Token (JWT) is used to control authentication and authorization. Hence, public can not access pages that can only be seen by authenticated users.

Connecting database and API on the server side should be as easy as possible.

To create API in no time and to configure database on server side, Strapi is used. 

To be able to create pages such as Login screen and Upload screen in this web application, Ant Design is involved in building this ID card repository. Nevertheless, placement of various web components can be a challenge in this web application as importing a web component sometimes needs extra configuration on style sheet.

Based on the introduction that has been described above, the following problem formulations are obtained :
\begin{enumerate}
    \item Does Strapi provide adequate requirement on the server side such as API creation, security, and  database configuration?
    \item Are required components available in Ant Design?
\end{enumerate}

\section{Literature Study}
Here are technology stacks that power the ID Card Repository. 

\subsection{ReactJS}
ReactJS is a JavaScript library which is deployed to develop reusable user interface (UI) components \autocite{aggarwal2018modern}. It is released by Facebook in 2013 and maintained by Facebook and a community consisting of developers and companies. ReactJS is optimized to provide speed, simplicity, and scalability \autocite{khuat2018developing}. The fundamental behind the React is the concept of virtual Document Object Manipulation (DOM) which React uses effectively. It renders sub-trees of nodes based upon state changes. In order to keep components up to date, React does the least amount of DOM manipulation possible. \autocite{kumar2016comparative} It utilizes JSX to create React elements. With power of JSX, there is no need to call React.createElement() each time when building virtual DOM. JSX is an optional HTML-like syntax that allows its users to create a virtual DOM tree without using the React.createElement() method. \autocite{fedosejev_react_2015} Other than JSX and virtual DOM, React has components which are categorized as stateful and stateless. Stateful components are used when data changes dynamically. State is a property of the React components that includes any data that a component needs. \autocite{vu2020building} Before React 16.8 was released, when data changes occur dynamically, a class-based component must be used. However, with the use of React hooks, a functional component can be used for it. React hooks is a mechanism for state management in a functional style that avoids the use of classes \autocite{madsen2020semantics}.

\subsection{Progressive Web Application}
Progressive Web Application (PWA) is a new set of standards advocated by the Google Web Fundamentals group to bridge a gap by introducing features such as offline support. PWA contributes toward the unification of the mobile experience where web applications can be installed and distributed without application marketplaces, work without Internet connectivity, receive push notifications, and look like regular applications. \autocite{biorn2017progressive} 

\subsection{Ant Design}
Ant Design is one of the most well-known ReactJS UI library that contains a set of high quality components and demos for building rich and interactive UI  \autocite{ant_design_ant_2020}. It is one of the most popular React UI library with more than 65000 stars on Github as well alongside Material UI, another famous React UI library. Even though it is said that Ant Design is a ReactJS UI library, Ant Design for VueJS and Angular are also available. Ant Design is applied on designing UI components.

\subsection{Strapi}
Strapi is a flexible and open-source headless content management system (CMS) that gives developers the freedom to choose their favorite tools and frameworks while also allowing editors to easily manage and distribute their content \autocite{strapi_welcome_2020}. Strapi can be used to create Application Programming Interface (API) fast and easily. It provides a user-friendly interface and allows users to create content-types which is similar with tables in databases without any technical knowledge \autocite{tanner2020implementation}. It has extensible plugin system including Admin Panel, Authentication \& Permission Management, Content Management and API Generator. \autocite{espinosa2020mobile}. It offers Restful API (REST API) by default. However, GraphQL API can be used alongside REST API by installing a GraphQL plugin. It is possible to install a GraphQL plugin either through UI or terminal/command prompt. With REST API, you might test your API using a third-party software, such as Postman. However, if you install GraphQL plugin, Strapi provides GraphQL Playground which helps a lot on testing your GraphQL API. It is accessible at '/graphql' endpoint. Strapi cares about database configuration as well. It can be used with various database management systems, such as MySQL, MongoDB, SQLite, and PostgreSQL. To use one of them, installing locally the one that will be used is a must. For instance, MongoDB is used to store data of ID Card repository. Therefore, installing MongoDB in the machine that will be used is a requirement.

\subsection{MongoDB}
MongoDB is one of the most popular document based database \autocite{gyHorodi2015comparative}. It is a very scalable NoSQL database management system that stores data in a BSON format, a dynamic schema document structured like Javascript Object Notation (JSON) \autocite{truica2013crud}.

\subsection{GraphQL}
GraphQL is a query language for APIs and a runtime for fulfilling those queries with your existing data. It provides clients the power to ask for exactly what they need and makes it easier to evolve APIs over time. \autocite{graphql_graphql_2020} Unlike Restful API that changes its version when it needs to upgrade, GraphQL API enables its users to add new fields and types to GraphQL API without impacting existing queries. Unlike REST API that has several endpoints, GraphQL API only utilizes a single endpoint which is '/graphql'. Moreover, GraphQL uses a schema to describe the organization of the data and declarative query language to allow clients to access data \autocite{hartig2018semantics}.

\subsection{Git}
Git is one of the most popular version control system. Distributed version control systems and their hosting services such as Github and Gitlab has revolutionized how developers collaborate by allowing them to freely exchange and integrate code changes in a peer-to-peer fashion \autocite{german2016continuously}. Git keeps track of code changes, so if there is a change, what has been modified can be tracked through commit messages and git status.

\section{Methodology}
Creating API is one of the most vital things in building this ID card repository on the server side. In this PWA, GraphQL API is utilized within Strapi. However, before creating GraphQL API, Strapi requires its users to choose a database management system then configure it easily.  

\begin{figure}
    \centering
    \includegraphics[scale=0.3]{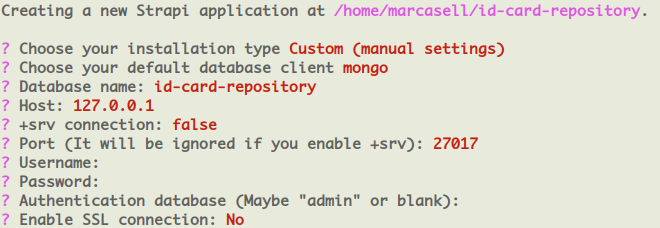}
    \caption{MongoDB configuration in Strapi.}
    \label{fig_config-mongo-INVERT}
\end{figure}

In setting up database management system through Strapi, develpoers are able to choose default configuration provided by Strapi as shown in \cref{fig_config-mongo-INVERT} or change it depends on what requirements of the application are.

\begin{figure}
    \centering
    \includegraphics[scale=0.2]{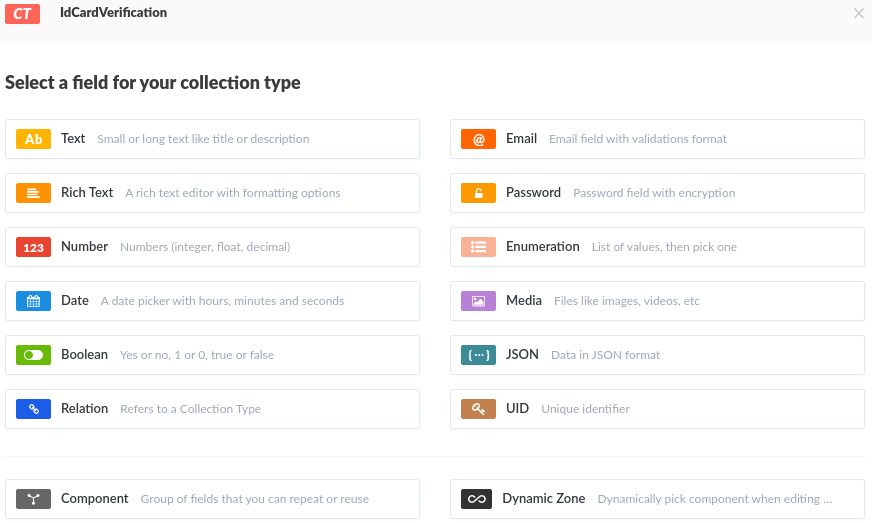}
    \caption{Various fields that are provided by Strapi}
    \label{fig_datatypes-in-Strapi}
\end{figure}

Creating a content-type is a step that is similar to determining fields and their data types in database management system. As shown in \cref{fig_datatypes-in-Strapi}, here are fields with their data types that can be chosen in Strapi :

\begin{itemize}
    \item Text
    \begin{itemize}
        \item Short Text
        \item Long Text
    \end{itemize}
    \item Rich Text
    \item Number
    \begin{itemize}
        \item integer
        \item big integer
        \item decimal
        \item float
    \end{itemize}
    \item Date
    \begin{itemize}
        \item date
        \item datetime
        \item time
    \end{itemize}
    \item Boolean
    \item Relation
    \item Email
    \item Password
    \item Enumeration
    \item Media
    \begin{itemize}
    \item Multiple Media
        \item Single Media
    \end{itemize}
    \item JSON
    \item UID
\end{itemize}

On the client side, Ant Design which has a lot of components can give a hand to style a web application. Not only components that usually can be seen in a web application like an input field and typography, it provides components to give spacing or margin as well. These are components of Ant Design that is used or planned to be utilized in the ID card repository on each page :

\begin{table}
\begin{center}
\caption{Ant Design Components in each screen}
\begin{tabular}{@{}ll@{}}
\toprule
Screen Name & Ant Design Component                                                                 \\ \midrule
Login       & \begin{tabular}[c]{@{}l@{}}- Form\\ - Input\\ - Button\\ - Space\end{tabular}        \\ \midrule
Register    & \begin{tabular}[c]{@{}l@{}}- Form\\ - Input\\ - Button\\ - Space\end{tabular}        \\ \midrule
Dashboard   & \begin{tabular}[c]{@{}l@{}}- Layout\\ - Table\\ - Breadcrumb\\ - Button\end{tabular} \\ \bottomrule
\end{tabular}
\end{center}
\end{table}

\begin{figure}
    \centering
    \includegraphics[scale=0.4]{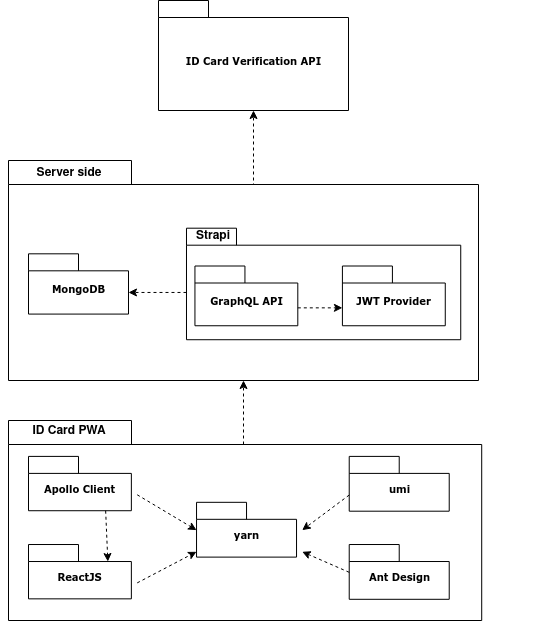}
    \caption{Dependencies of the PWA}
    \label{fig_UML}
\end{figure}

\cref{fig_UML} illustrates dependency graphs in ID card repository PWA. Basically there are three components which are ID Card PWA that acts as a client side, Server side that generates GraphQL API and JWT provider within Strapi that is connected with MongoDB, and the last one is ID Card Verification API whose function is to verify ID Cards which are uploaded to ID card repository. On the client side, everything is dependent on yarn as a package manager. On the server side, JWT provider gives a JWT as one of the authentication response alongside username and password that are fetched from MongoDB.

\begin{figure}
    \centering
    \includegraphics[scale=0.4]{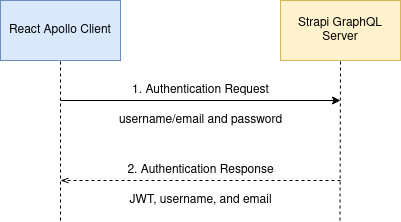}
    \caption{Authentication Flow of the ID Card Repository}
    \label{fig_authentication-flow}
\end{figure}

\cref{fig_authentication-flow} describes data flows on authentication process. First, React Apollo client will send an authentication request which contains username or email and password. Then Strapi GraphQL server will send back JWT, username, and email of the account as an authentication response.

\section{Evaluation Methodology}
Strapi provides GraphQL playground which is used to test GraphQL API. To evaluate GraphQL API, both registration and login queries were tested. Below is a query that register a user with GraphQL API :

\begin{lstlisting}
mutation createUser(
 $input: createUserInput
) {
 createUser(
  input: $input
 ) {
    user {
      username
      email
    }
  }
}
\end{lstlisting}

To login successfully, one must retrieve a JWT that will be stored in local storage. Below is a GraphQL query to obtain a JWT :
\begin{lstlisting}
mutation Login(
 $input: UsersPermissionsLoginInput!
 ) {
  login(
   input: $input
  ) {
    jwt
    user {
      username
      email
    }
  }
}
\end{lstlisting}

\section{Evaluation Result}
Strapi can be utilized to create API fast. Soon after selecting fields and saving the content type, the API will be created. Table 1 below is the fields in a content type of Strapi GraphQL API with their data types :

\begin{table}[htbp]
\begin{center}
\caption{Fields and Data Types in ID Card Repository PWA}
\begin{tabular}{@{}ll@{}}
\toprule
\multicolumn{1}{c}{\textbf{Field}} & \multicolumn{1}{c}{\textbf{Data Type}} \\ \midrule
kind                               & enumeration                            \\
identifier                         & short text                             \\
name                               & short text                             \\
birthPlace                         & short text                             \\
birthDate                          & date                                   \\
gender                             & enumeration                            \\
bloodType                          & enumeration                            \\
address                            & long text                              \\
religion                           & short text                             \\
marriageStatus                     & enumeration                            \\
occupation                         & short text                             \\
nationalityCode                    & short text                             \\
expiryDate                         & date                                   \\
facePhoto                          & single media                           \\
cardImage                          & single media                           \\
personWithCardPhoto                & single media                           \\
issuerCountryCode                  & short text                             \\
issuedDate                         & date                                   \\
faceTop                            & integer                                \\
faceLeft                           & integer                                \\
faceWidth                          & integer                                \\
faceHeight                         & integer                                \\
statusCode                         & enumeration                            \\
uploadedAt                         & datetime                               \\
extractedAt                        & datetime                               \\
verifiedAt                         & datetime                               \\
issuerProvince                     & short 

text                             \\
issuerCity                         & short text                             \\
uploaderId                         & short text                             \\ \bottomrule
\end{tabular}
\end{center}

\end{table}

\begin{figure}
    \centering
    \includegraphics[scale=0.3]{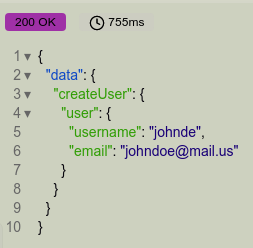}
    \caption{Strapi GraphQL server Registration response}
    \label{fig_api-1}
\end{figure}

In \cref{fig_api-1}, Strapi GraphQL server responds to createUser mutation in 755ms.  

\begin{figure}
    \centering
    \includegraphics[scale=0.3]{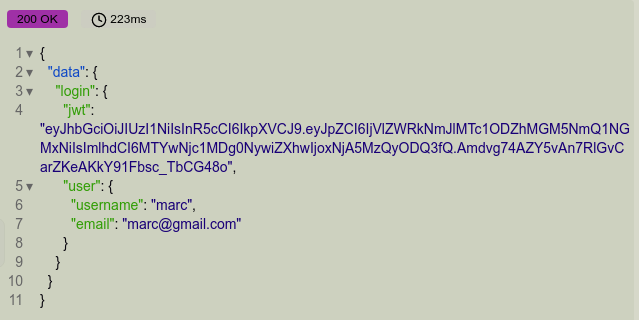}
    \caption{Strapi GraphQL server Authentication Response}
    \label{fig_api-2}
\end{figure}

\cref{fig_api-2} shows Strapi GraphQL server gives response to Login mutation in 223ms.

\begin{figure}
    \centering
    \includegraphics[scale=0.44]{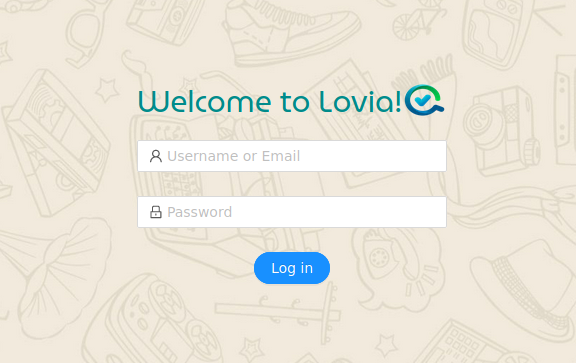}
    \caption{Login screen build with Ant Design}
    \label{fig_login-1}
\end{figure}

Having Strapi run on the server side, on the client side, The styling was done with Ant Design, including its web components and icons as shown in \cref{fig_login-1}.

\section{Conclusion}
Strapi comes up with required fields and data types to create ID card repository. It also helps developers on API creation quickly and easily. GraphQL API and mutation queries that were tested with GraphQL playground in Strapi worked well as well. Ant Design that has a lot of web components enables styling to be done easily alongside ReactJS and Apollo Client on the client side.  

For the future development, connecting to ID card verification API, creating registration, upload, and dashboard screens, making UI responsive, and implementing a JWT to control authentication and authorization flow are planned.

\section*{Acknowledgment}
First, we would like to thank Lovia (https://about.lovia.life), who supported us to finish this research paper. Next, we would like to thank to our parents, who supported us with love and understanding. Without them, we could never have reached this current level of success. 

\printbibliography[]
\vspace{12pt}

\end{document}